\documentclass[10pt,conference]{IEEEtran}

\usepackage{cite}
\usepackage{amsmath,amssymb,amsfonts}
\usepackage{algorithmic}
\usepackage{graphicx}
\usepackage{textcomp}
\usepackage{xcolor}
\def\BibTeX{{\rm B\kern-.05em{\sc i\kern-.025em b}\kern-.08em
    T\kern-.1667em\lower.7ex\hbox{E}\kern-.125emX}}

\usepackage[utf8]{inputenc} 
\usepackage[T1]{fontenc}    
\usepackage[hidelinks]{hyperref}
\usepackage{url}            

\usepackage{booktabs}       
\usepackage{nicefrac}       
\usepackage{microtype}      
\usepackage{tabularx} 
\usepackage{listings}       
\usepackage{multirow}
\usepackage{array}
\usepackage{makecell}
\usepackage[most]{tcolorbox}

\newtcolorbox{agentpanel}[1]{
  colback=gray!5,
  colframe=black!60,
  boxrule=0.5pt,
  arc=1pt,
  left=6pt,right=6pt,top=6pt,bottom=6pt,
  title={#1},
  fonttitle=\bfseries,
  before upper={\raggedright\setlength{\parindent}{0pt}}
}

\lstdefinestyle{skillstruct}{
  basicstyle=\ttfamily\footnotesize,
  columns=fullflexible,
  keepspaces=true,
  showstringspaces=false,
  breaklines=true,
  breakatwhitespace=false,
  frame=single,
  framerule=0.3pt,
  xleftmargin=2pt,
  xrightmargin=2pt
}

\usepackage[ruled,lined,linesnumbered,vlined]{algorithm2e}
\newcommand{\paneldivider}{\par\smallskip\hrule\smallskip}

\usepackage{caption}
\begin{document}

\title{Do Skill Descriptions Tell the Truth? Detecting Undisclosed Security Behaviors in Code-Backed LLM Skills}


\author{
\IEEEauthorblockN{Wenhui He, Yue Li, Bang Fu, Huan Xing, Xing Fan, ZeHua Zhang, Baoning Niu}
\IEEEauthorblockA{
Skill Security Team\\
}
}


\maketitle

\begin{abstract}
  Programmatic skills in LLM ecosystems consist of a natural-language description and executable implementation files. Users and LLMs rely on the description to understand the skill's scope. However, the implementation may perform security-relevant operations, such as credential access, network communication, or command execution, that the description does not state. We study this description--implementation inconsistency by asking whether the implementation stays within the security-relevant scope declared in the description. We manually analyze 920 real-world programmatic skills and construct an 11-category security property taxonomy. Based on this taxonomy, we build \textsc{SkillScope}, which constructs source-level security property graphs (SPGs) from implementations and performs LLM-assisted consistency checking. SPG nodes retain source-level code patterns rather than abstract taxonomy labels, preserving fine-grained evidence for checking. On 4,556 programmatic skills with double-blind human review, \textsc{SkillScope} achieves a precision of 84.8\% and a recall of 96.5\% for identifying inconsistency. Confirmed inconsistency affects 9.4\% of skills, while cases of coarser description, in which implementation details remain within the declared scope, account for 24.3\%. Ablation experiments confirm that both the SPG and the taxonomy contribute: removing the taxonomy reduces precision from 87.8\% to 72.3\%, while removing the SPG reduces recall from 94.7\% to 79.0\%.
\end{abstract}


\section{Introduction}
\label{sec:introduction}
LLM skill ecosystems are expanding, and reusable skills have become a common way to package task-specific capabilities~\cite{anthropic_skills,github_about_skills,openai_skills}. Some skills include executable code and a natural-language description file. Different platforms use different file structures, but they share a common pattern: a description file, often named \texttt{SKILL.md}, specifies the skill's intended functionality in natural language. Some skills also include one or more implementation files that define executable behavior~\cite{anthropic_skills,github_create_skills}. In this work, we refer to skills with such implementation files as \textbf{programmatic skills}, distinguishing them from prompt-only skills that rely only on natural-language instructions. Throughout this paper, we use \emph{description} to refer to the entire \texttt{SKILL.md} file, including any declarative metadata header it provides (with fields such as \texttt{description:} and \texttt{instructions:}) and any free-form natural-language content. We use \emph{implementation} to refer to the executable files shipped with the skill. Unless otherwise stated, ``description'' in this paper never denotes the \texttt{description:} metadata field alone.

Users and LLMs rely on the description to understand the skill's scope before use, but the implementation can invoke external tools, access local files, communicate over the network, or handle credentials~\cite{anthropic_overview,anthropic_security,github_create_skills}. If the implementation performs security-relevant operations that the description does not state, users and LLMs may unknowingly invoke capabilities beyond what they intended. This raises a question: does the implementation stay within the security-relevant scope declared in the description?

We define \emph{description--implementation inconsistency} as cases where the implementation's security-relevant behaviors exceed the scope declared in the description. Such inconsistency takes two forms. \emph{Undeclared behavior} occurs when the implementation contains a security-relevant operation not covered by the description. \emph{Undeclared flow} occurs when the implementation contains a data or control flow path among security-relevant operations that the description does not reflect. Not every description--implementation difference constitutes inconsistency. When the implementation operates within the capabilities already covered by the description but at a finer level of detail, we characterize the relationship as \emph{coarser description}: implementation details remain within the declared scope but are more specific than what the description states.

To study this problem, we manually analyze 920 real-world programmatic skills to derive a security property taxonomy of 11 first-level categories, producing a reference annotation set for description-side security properties. We then build \textsc{SkillScope}, which constructs code-side security property graphs (SPGs) from skill implementations and performs LLM-assisted consistency checking using the taxonomy as a scope constraint. SPG nodes retain source-level operation patterns rather than abstract taxonomy labels, preserving fine-grained evidence for checking. On the full dataset of 4,556 programmatic skills, \textsc{SkillScope} achieves a precision of 84.8\% and a recall of 96.5\% for identifying description--implementation inconsistency. Ablation experiments on a 300-skill subset confirm that both the SPG and the taxonomy contribute: removing the taxonomy reduces precision from 87.8\% to 72.3\%, while removing the SPG reduces recall from 94.7\% to 79.0\%.

We make four contributions:
\begin{itemize}
    \item A security property taxonomy covering 11 categories of security-relevant behaviors, derived from manual analysis of 920 real-world programmatic skills, together with a reference annotation set for description-side security properties.
    \item \textsc{SkillScope}, a tool that constructs source-level security property graphs from skill implementations and performs consistency checking against the description, achieving a precision of 84.8\% and a recall of 96.5\% on 4,556 programmatic skills.
    \item An empirical analysis of 4,556 programmatic skills with double-blind human review of the full dataset, revealing that 9.4\% exhibit confirmed inconsistency and 24.3\% exhibit coarser description, and identifying the dominant inconsistency and granularity-mismatch patterns.
    \item Ablation experiments and cross-model comparisons that quantify the contribution of each pipeline component.
\end{itemize}

\section{Background and Motivation}
\label{sec:background}


Figure~\ref{fig:code_backed_skill} shows the structure of a programmatic skill. The description and implementation are located in the same skill directory but serve distinct roles: the description declares what the skill does, while the implementation defines how the skill executes~\cite{anthropic_skills,github_add_skills}. However, no mechanism enforces that the description covers all behaviors in the implementation, so the description may not fully reflect the security-relevant operations that the code actually performs.

\subsection{Motivating Example}

Consider a skill whose description states: ``Read target files and run a fixed analysis workflow.'' The implementation, however, contains \texttt{os.getenv(`API\_KEY')} followed by \texttt{requests.post(endpoint, headers=\{...\})}. Two forms of inconsistency are present. First, an undeclared behavior: the implementation accesses a credential not mentioned in the description. Second, an undeclared flow: a local secret is read and transmitted to an external endpoint, forming a local-to-external data flow path absent from the description. Reading the description alone would not indicate either operation.

In contrast, if a description states ``read local files and produce a summary report,'' and the implementation writes the report to a specific path such as \texttt{output/report.json}, this difference reflects finer detail within the scope already covered by the description. We classify such cases as coarser description rather than a security-boundary violation.

\subsection{Threat Model}

We consider settings where users and LLMs derive their security expectations from the description before invocation, while the implementation governs the actual runtime behavior. A skill author, whether through oversight or intent, may omit or understate security-relevant behaviors or flows in the description.

We assume the following. The skill author controls the skill's content, including both \texttt{SKILL.md} and the implementation files. Users and LLMs treat the description as the security boundary and do not independently audit the implementation before use. The platform does not perform automated consistency checking at the time of this study.

Our goal is to assess whether the implementation exceeds the security boundary declared in the description. We take the description as the reference boundary and check for undeclared behaviors or undeclared flows beyond that boundary. We do not attribute intent or determine whether a mismatch led to exploitation.

\begin{figure}[t]
\centering
\begin{minipage}{0.95\linewidth}
\begin{lstlisting}[style=skillstruct]
programmatic skill/
+-- SKILL.md                  # Description layer
+-- scripts/                  # Implementation layer
|   +-- format_code.py        # Example implementation file

# SKILL.md
---
description: Format source code files using project conventions
instructions: Read source files, apply formatting, write results
---
Apply the configured formatting style to each source file.

# format_code.py
def format_project(src_dir):
    files = discover_sources(src_dir)
    formatted = apply_style(files)
    write_formatted(formatted)
\end{lstlisting}
\end{minipage}
\caption{Simplified structure of a programmatic skill. The \texttt{SKILL.md} file forms the description layer, while scripts or other implementation files define executable behavior.}
\label{fig:code_backed_skill}
\end{figure}

\section{Security Property Investigation}
\label{sec:sec_pro}

\begin{figure*}[htbp]
    \centering
    \includegraphics[width=0.95\linewidth]{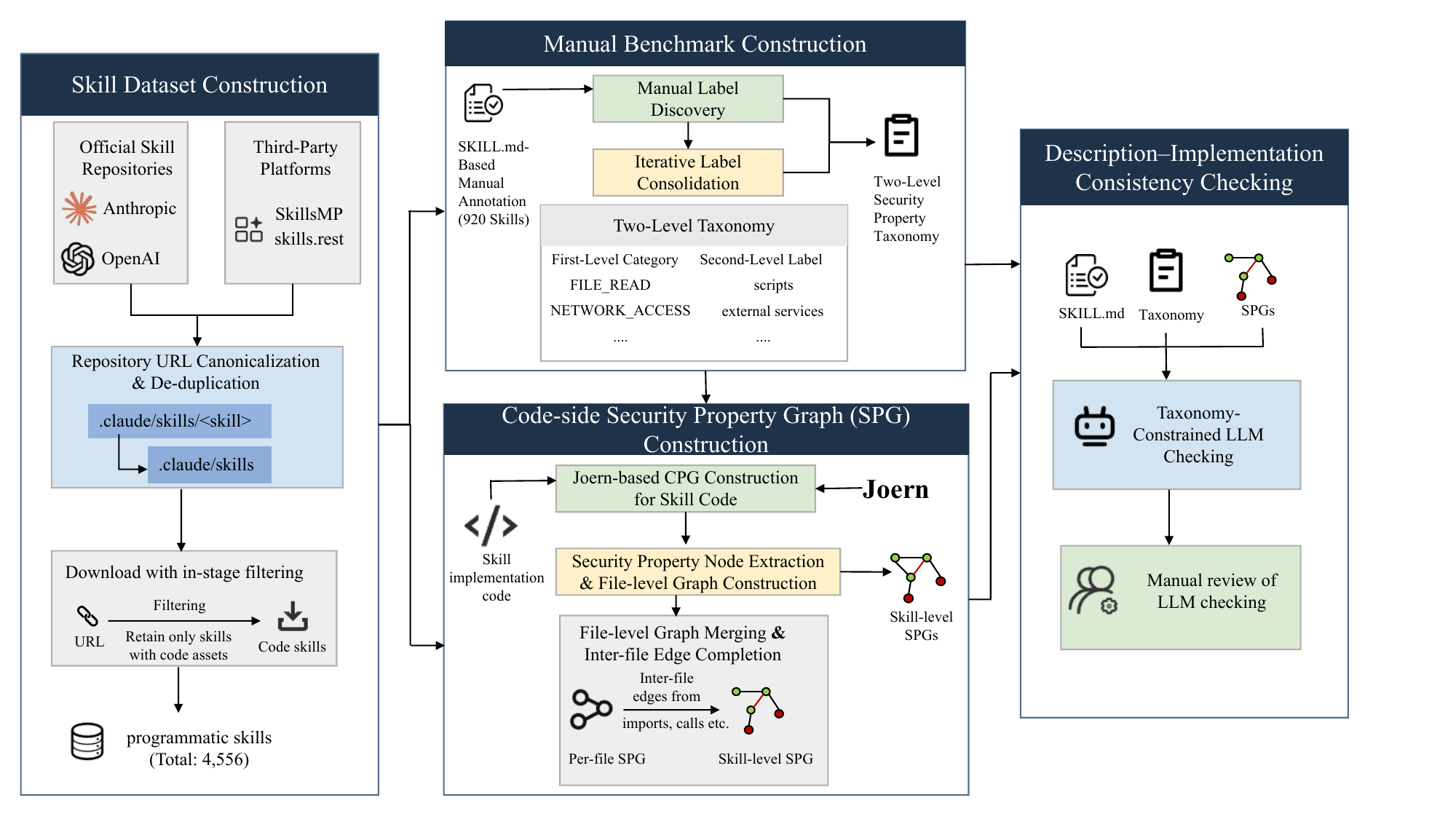}
    \caption{Overview of the analysis pipeline. The pipeline includes dataset construction, taxonomy derivation and manual annotation from \texttt{SKILL.md}, code-side security property graph construction, and description--implementation consistency checking.}
    \label{fig:overall_pipeline}
\end{figure*}

Detecting inconsistency at scale requires a concrete definition of what constitutes a security-relevant behavior and a dataset of real-world skills to analyze. Figure~\ref{fig:overall_pipeline} shows the overall workflow. This section presents the construction of the dataset and the security property taxonomy. The remaining components, SPG construction and consistency checking, are presented in Section~\ref{sec:skillscope}.

\subsection{Dataset Construction}
\label{subsec:dataset}

The dataset is based on a public snapshot collected as of March 5, 2026. We collect skill entries from four sources: two first-party sources (Anthropic official skills and OpenAI) and two community aggregators (SkillsMP and skills.rest)~\cite{anthropic_skills,openai_skills,skillsmp_home,skillsrest_home}. Across these sources we obtain 22,515 raw skill items, each linked to a public GitHub repository or directory.

\paragraph{URL Normalization and Deduplication.}
Links to the same skill may take different forms across sources, and multiple links may also refer to the same skill directory in the same repository. To unify these links, we parse each GitHub URL of the form \nolinkurl{https://github.com/<owner>/<repo>/tree/<branch>/<path>} into four components: \texttt{owner}, \texttt{repo}, \texttt{branch}, and \texttt{subpath}, where \texttt{subpath} is the repository-relative path after removing the repository and branch prefix. We then normalize \texttt{subpath} to account for source-specific differences in directory layout, for example mapping \texttt{.claude/skills/<skill>} to \texttt{.claude/skills}, and retaining the nearest \texttt{.../skills} prefix for nested paths. After normalization, we use \texttt{owner/repo@branch:normalized\_subpath} as the grouping key and merge links that map to the same download target. This deduplication reduces the 22,515 raw items to 11,049 unique download targets.

\paragraph{Content-Based Filtering.}
We then apply content-based filtering. Because the later analysis requires both a description and implementation files, we retain a skill directory only if it contains a \texttt{SKILL.md} file and at least one analyzable implementation file, identified by suffix (\texttt{.py}, \texttt{.js}, \texttt{.ts}, or \texttt{.go}). Skills that contain only \texttt{SKILL.md}, documentation files, images, data files, or other non-code files are excluded. Of the 11,049 unique download targets, 4,556 satisfy these criteria and form the final dataset of programmatic skills. Of these 4,556 skills, 17.6\% come from the two first-party sources (Anthropic and OpenAI) and the remaining 82.4\% from the two community aggregators. A single skill may include files in more than one of the four analyzable languages, so we do not report a strict per-language partition.

\subsection{Security Property Taxonomy}
\label{subsec:taxonomy}
We construct a two-level security property taxonomy that defines the scope of security-relevant behaviors considered in this work. The taxonomy contains 11 first-level categories and 32 second-level labels. The first-level categories cover security-relevant behavior classes: file reads, file writes, command execution, network access, external API usage, secret access, dependency modification, system permission access, security control, observability, and infrastructure. The second-level labels specify the target of each behavior, such as API keys, session data, and identity information under \texttt{SECRET\_ACCESS}. To reduce annotation ambiguity, the taxonomy also provides operational descriptions and boundary rules for overlapping categories, such as \texttt{NETWORK\_ACCESS} and \texttt{EXTERNAL\_API}. Table~\ref{tab:two_level_taxonomy} shows the full taxonomy.

\paragraph{Annotation Workflow}
We derive the taxonomy through manual analysis of 920 programmatic skills, sampled from the 4,556-skill dataset in proportion to the source distribution; these 920 annotated skills therefore form a subset of the evaluation set used in Section~\ref{sec:evaluation}. The annotation labels which description-side security-property categories each \texttt{SKILL.md} explicitly declares; it does not label inconsistency itself, which is judged only during the evaluation review.

We deliberately avoid LLM-assisted labeling at this stage: LLM-generated labels may infer capabilities that are not explicitly stated and may apply unstable boundaries to similar statements, merging them in some cases and splitting them in others. Two authors therefore independently annotate the 920 skills, considering only security properties explicitly stated in \texttt{SKILL.md} without consulting implementation files or inferring unstated capabilities. The annotation proceeds in two stages. In the \emph{pilot stage}, both annotators independently annotate the same 120-skill subset; we begin with official samples because their descriptions and implementation organization are more regular, which helps stabilize the initial criteria, but the pilot subset itself is not limited to official samples. After the pilot, the annotators compare results, discuss disagreements, and align criteria and label boundaries. In the \emph{formal stage}, both annotators independently annotate the remaining 800 skills under the aligned criteria. On the pilot subset, exact-match agreement before adjudication is 90.0\% (108/120) at the first level and 78.3\% (94/120) at the second level; Table~\ref{tab:benchmark_summary} summarizes dataset construction and the annotation setting.

\paragraph{Conflict Resolution}
For samples with disagreements or boundary uncertainty, the annotators first discuss the case based only on the original description text, the current taxonomy, and the annotation criteria; implementation files are not consulted at this stage. They focus on two questions: (i) does the statement explicitly state a description-side security property, and (ii) can the statement be covered by an existing property label. If the annotators still cannot reach agreement, the case is submitted to a designated adjudicator (the first author), whose decision determines the final label.

\paragraph{Taxonomy Refinement}
The taxonomy can be refined during annotation, but we do not update it based on a single new case. Such cases are first recorded for later review by the two annotators and the adjudicator. A label definition is revised, or a new label is added, only when the phenomenon cannot be covered by existing property labels and recurs across multiple samples with a stable boundary. After each update, the team applies the revised criteria consistently and, when necessary, revisits previously annotated samples.

The taxonomy serves three roles in our pipeline: (1) it defines the label space for annotating description-side security properties; (2) it scopes code-side SPG node localization to the same set of categories; and (3) it provides a common reference for consistency checking by both \textsc{SkillScope} and the human reviewers, so that their judgments are directly comparable.

\begin{table*}[t]
\caption{Two-level description-side security property taxonomy.}
\label{tab:two_level_taxonomy}
\centering
\scriptsize
\setlength{\tabcolsep}{4pt}
\renewcommand{\arraystretch}{1.05}
\begin{tabularx}{\textwidth}{
>{\raggedright\arraybackslash}p{0.18\textwidth}
>{\raggedright\arraybackslash}p{0.16\textwidth}
>{\raggedright\arraybackslash}X}
\toprule
\textbf{First-level label} & \textbf{Second-level label} & \textbf{Operational description} \\
\midrule

\multirow{4}{0.18\textwidth}{FILE\_READ}
& FR-SCRIPT & Read local scripts, source files, or implementation files. \\
& FR-REF & Read local reference documents, guides, templates, or instruction files. \\
& FR-DATA & Read local input data, project files, datasets, or media files. \\
& FR-CONFIG & Read local configuration, state, database, session, environment, or queue files. \\
\midrule

\multirow{4}{0.18\textwidth}{FILE\_WRITE}
& FW-OUTPUT & Write local output artifacts, such as reports or generated files. \\
& FW-STATE & Write local state data, such as databases, caches, logs, metrics, queues, or other persistent state files. \\
& FW-CONFIG & Write local configuration, environment, authentication, session, or browser-state files. \\
& FW-STRUCTURE & Create directories, project structures, template instances, or scaffold files. \\
\midrule

\multirow{2}{0.18\textwidth}{SYSTEM\_COMMAND}
& SC-PY & Execute local Python scripts or skill-provided scripts. \\
& SC-CLI & Execute shell, npm, npx, pip, docker, git, chmod, sqlite3, build, or test commands. \\
\midrule

\multirow{3}{0.18\textwidth}{NETWORK\_ACCESS}
& NA-WEB & Access websites, web pages, browser content, web search, or localhost services. \\
& NA-SERVICE & Access remote platforms, cloud services, model services, or SaaS systems. \\
& NA-DOWNLOAD & Clone, pull, or download remote repositories, images, models, datasets, or documents. \\
\midrule

\multirow{3}{0.18\textwidth}{EXTERNAL\_API}
& EA-DATA & Access remote databases, metadata services, or search APIs. \\
& EA-PLATFORM & Access remote platform APIs or business-system APIs. \\
& EA-AI & Access remote AI, model, inference, or search APIs. \\
\midrule

\multirow{3}{0.18\textwidth}{SECRET\_ACCESS}
& SA-KEY & Access API keys, passwords, OAuth tokens, access tokens, or similar secrets. \\
& SA-SESSION & Access cookies, browser sessions, authentication state, or similar session data. \\
& SA-ID & Access identity-related information, such as email addresses, mailto fields, or user identifiers. \\
\midrule

\multirow{3}{0.18\textwidth}{\makecell[l]{DEPENDENCY\_\\MODIFICATION}}
& DM-PKG & Install or update package dependencies through pip, uv, npm, or similar package managers. \\
& DM-SYS & Install system-level dependencies, browsers, OCR tools, Playwright, TeX, or similar system components. \\
& DM-ENV & Create or modify virtual environments, Conda environments, containers, or other execution environments. \\
\midrule

\multirow{5}{0.18\textwidth}{\makecell[l]{SYSTEM\_PERMISSION\_\\ACCESS}}
& SPA-RESOURCE & Access system resources or related permissions. \\
& SPA-IAM & Manage IAM roles or permissions. \\
& SPA-VALIDATION & Perform security validation or permission checks. \\
& SPA-ENFORCEMENT & Enforce access restrictions or permission limits, including access-control and compliance checks. \\
& SPA-SECURITY & Perform security-related operations. \\
\midrule

\multirow{3}{0.18\textwidth}{SECURITY\_CONTROL}
& SEC-VALIDATION & Validate inputs. \\
& SEC-QUERY & Perform parameterized queries. \\
& SEC-RATE & Apply rate limiting. \\
\midrule

OBSERVABILITY
& OBS-LOG & Produce structured logs. \\
\midrule

INFRASTRUCTURE
& INF-HEALTH & Expose health-check endpoints. \\
\bottomrule
\end{tabularx}
\end{table*}

\begin{table}[t]
\caption{Summary of dataset construction, manual annotation setting, and exact-match agreement.}
\label{tab:benchmark_summary}
\centering
\small
\setlength{\tabcolsep}{4pt}
\renewcommand{\arraystretch}{1.05}
\begin{tabularx}{\columnwidth}{>{\raggedright\arraybackslash}p{0.42\columnwidth} >{\raggedright\arraybackslash}X}
\toprule
\textbf{Item} & \textbf{Value} \\
\midrule
Public snapshot date & March 5, 2026 \\
Final dataset & 4,556 programmatic skills \\
Manually annotated subset & 920 skills \\
Pilot annotation (double-annotated) & 120 skills \\
Formal annotation (double-annotated) & 800 skills \\
Annotators & 2 authors \\
Annotation source & \texttt{SKILL.md} only \\
Annotation unit & One programmatic skill \\
Adjudication & First author \\
Pilot first-level exact-match agreement & 108/120 (90.0\%) \\
Pilot second-level exact-match agreement & 94/120 (78.3\%) \\
\bottomrule
\end{tabularx}
\end{table}

\section{SkillScope}
\label{sec:skillscope}

Building on the taxonomy and reference annotation set from Section~\ref{sec:sec_pro}, \textsc{SkillScope} constructs a code-side security property graph (SPG) from the implementation and checks whether the implementation exceeds the scope declared in the description.

Formally, for an implementation file \(f\), the file-level SPG is \(G_f = (V_f, E_f)\), where \(V_f\) is the set of security property nodes identified from security-relevant implementation sites in \(f\), and \(E_f \subseteq V_f \times V_f\) is the set of directed intra-file edges among these nodes, discovered through breadth-first search (BFS) reachability analysis on the intermediate representation. Each node \(v \in V_f\) corresponds to one security-relevant implementation location and is associated with its source location and source-level operation pattern. The graph contains only security property nodes and excludes program nodes that are unrelated to the security properties considered in this work. The BFS traversal follows only data-flow edges (from the program dependence graph, PDG) and control-flow edges (from the control-flow graph, CFG); abstract syntax tree (AST) structural edges are excluded to avoid spurious connections through syntactic containment. For a skill \(s\) with implementation files \(\mathcal{F}_s\), the skill-level SPG is \(G_s = (V_s, E_s)\) with \(V_s = \bigcup_{f \in \mathcal{F}_s} V_f\) and \(E_s = \bigcup_{f \in \mathcal{F}_s} E_f \cup E_s^{\text{cross}}\), where \(E_s^{\text{cross}}\) denotes cross-file edges connecting existing SPG nodes across files without introducing new nodes.

\subsection{Code-Side SPG Construction}

\textsc{SkillScope} constructs an SPG from the implementation files of each skill following Algorithm~\ref{alg:spg_construction}. We first export a Code Property Graph (CPG)~\cite{yamaguchi2014modeling} of the skill folder using Joern~\cite{joern_docs} as the intermediate representation (Line 3). Joern's CPG integrates AST, CFG, and PDG into a unified structure, providing both structural and flow-level information. SPG nodes are identified using taxonomy-derived localization rules, and edges are discovered through reachability analysis on the CPG. Joern is used only as the extraction source; after SPG extraction, the CPG is discarded. Because the localization rules are derived from the taxonomy rather than from the individual skill's description, \textsc{SkillScope} can detect security-relevant behaviors in the implementation even when they are not mentioned in the description.

\paragraph{Node Localization}
For each taxonomy category, we define a set of keyword-based localization rules that match security-relevant call sites in the source code. The rules operationalize the behavior classes defined by the taxonomy into source-level localization criteria: the taxonomy specifies what kinds of security-relevant behaviors should be recognized, while the localization rules specify what kinds of implementation sites are treated as evidence of such behaviors in code. For example, \texttt{FILE\_READ} is matched by patterns such as \texttt{open(} and \texttt{readFile(}; \texttt{FILE\_WRITE} by \texttt{f.write(} and \texttt{createWriteStream(}; \texttt{SECRET\_ACCESS} by \texttt{os.getenv(} and \texttt{process.env.}; and \texttt{NETWORK\_ACCESS} or \texttt{EXTERNAL\_API} by \texttt{requests.post(}, \texttt{fetch(}, and SDK-specific call patterns. Each matched site becomes an SPG node (Line 6); the rule set covers all 11 taxonomy categories, and the full pattern list is provided in our supplementary artifact.

\paragraph{Node Representation}
SPG nodes retain the source-level operation pattern at each matched site rather than abstracting it to a taxonomy label. For example, a node records \texttt{os.getenv(`API\_KEY')} or \texttt{requests.post(endpoint, headers=\{...\})}, preserving the specific call, argument, and target. We make this design choice because taxonomy labels are too coarse for accurate consistency checking: reading a private key file and reading a local JSON data file both map to \texttt{FILE\_READ}, but their security implications differ substantially. Retaining source-level patterns allows the LLM to compare the concrete operation against the description rather than relying on an abstraction that omits security-relevant details.

\paragraph{Edge Discovery}
Given the set of SPG nodes \(V_f\) in a file \(f\), we discover intra-file edges (Line 7) by checking, for each pair of nodes in \(V_f\), whether a data-flow or control-flow path connects them in the CPG. We perform BFS on the CPG starting from one node; if another node in \(V_f\) is reached, we add a directed edge from the starting node to the reached node. The BFS traversal may pass through intermediate non-security program nodes that are not themselves part of \(V_f\); these intermediate nodes are not added to the SPG, but the reachability they establish produces an SPG edge. This approach captures indirect flows. For example, a credential read by \texttt{os.getenv()} may reach \texttt{requests.post()} through several intermediate variable assignments. Although there is no direct CPG edge between the two security nodes, BFS discovers the path and adds an SPG edge. All reachable pairs in \(V_f \times V_f\) contribute edges, forming the intra-file edge set \(E_f\).

\paragraph{Skill-Level Merging}
After each file is processed, its SPG is accumulated into the skill-level \(V_s, E_s\) (Lines 8--11). The algorithm then iterates over distinct file pairs and adds cross-file edges (Lines 12--14) based on explicit dependency evidence: import or require relations, resolved cross-file call references, and path-level dependencies. Cross-file edges connect existing SPG nodes across files without introducing new nodes; the resulting \(G_s = (V_s, E_s)\) is returned at Line 15.

As an example of the resulting skill-level representation, the SPG for a PDF-processing skill with 8 implementation files contains \(|V_s|=250\) security-relevant nodes and \(|E_s|=301\) directed edges. The nodes cover 18 distinct source-level operation patterns, including \texttt{open(\dots)}, \texttt{sys.argv[i]}, and \texttt{json.load(\dots)}. This illustrates that the aggregation summarizes multi-file implementations through security-relevant operations and their connections, rather than preserving the full program graph.

\begin{algorithm}[t]
\footnotesize
\DontPrintSemicolon
\SetCommentSty{algcommentfont}
\SetKwInOut{Input}{Input}
\SetKwInOut{Output}{Output}

\Input{
    $p_s$: folder path of a programmatic skill. \\
    $\mathcal{R}$: property localization criteria derived from the taxonomy. \\
    $\mathcal{C}$: cross-file edge completion criteria.
}
\Output{
    $G_s = (V_s, E_s)$: skill-level code-side SPG.
}

$V_s \gets \emptyset,\; E_s \gets \emptyset$ \\
$S \gets \emptyset$ \\
$CPG_s \gets \mathtt{ExportCPG}(p_s)$ \tcp*{Joern exports CPG; used only as extraction source for SPG}

\tcp{Phase I: File-level SPG extraction}
\ForEach{implementation file $f$ in $p_s$}{
    $CPG_f \gets \mathtt{GetFileSubgraph}(CPG_s, f)$ \\
    $V_f \gets \mathtt{LocateSecurityNodes}(CPG_f, \mathcal{R})$ \tcp*{Taxonomy-derived node localization}
    $E_f \gets \mathtt{DiscoverEdgesByBFS}(CPG_f, V_f)$ \tcp*{BFS reachability between SPG nodes}
    $G_f \gets (V_f, E_f)$ \\
    $S \gets S \cup \{(f, G_f)\}$ \\
    $V_s \gets V_s \cup V_f$ \\
    $E_s \gets E_s \cup E_f$
}

\tcp{Phase II: Skill-level merging and cross-file edge completion}
\ForEach{pair of distinct entries $((f_i, G_{f_i}), (f_j, G_{f_j}))$ in $S$}{
    $E_{cross} \gets \mathtt{CompleteCrossFileEdges}(G_{f_i}, G_{f_j}, \mathcal{C})$ \\
    $E_s \gets E_s \cup E_{cross}$
}

\Return{$G_s = (V_s, E_s)$}
\caption{Code-side SPG construction for a programmatic skill}
\label{alg:spg_construction}
\end{algorithm}

\subsection{Description--Implementation Consistency Checking}

After constructing the SPG, \textsc{SkillScope} checks whether the implementation exceeds the scope declared in the description. We keep the description in its original text form and represent only the implementation as an SPG, because the description is typically more abstract and forcing it into a graph would introduce unstable node and edge alignment. The LLM takes three inputs: the description (\texttt{SKILL.md}), the serialized SPG (\texttt{code\_graph\_json}), and the security property taxonomy.

\paragraph{Checking Criteria.}
The checking produces two outputs in a single pass. The first is whether description--implementation inconsistency exists. We define inconsistency based on two conditions:
\begin{itemize}
    \item \textbf{(C1)} The implementation contains one or more security-relevant objects that are not covered by any capability declared in the description. These objects include credential types, external entities, system permission scopes, or persistence targets that are neither mentioned nor implied by any declared capability.
    \item \textbf{(C2)} The implementation contains a data or control flow path that crosses security domain boundaries (e.g., local secret to external endpoint, user input to system command execution), and the description does not cover this path.
\end{itemize}
If any node or edge in the SPG satisfies C1 or C2, the skill is classified as exhibiting inconsistency. The second output is whether the skill exhibits coarser description. This applies when the undeclared details remain within a capability already declared in the description and do not satisfy C1 or C2.

\paragraph{Prompt Design}
Figure~\ref{fig:consistency_checking_prompt} shows the prompt template. The prompt structures the checking as a sequential process. The LLM first extracts the security semantics explicitly declared in the description. It then examines each SPG node against the declared semantics, applying C1 to identify undeclared security-relevant objects. Next, it examines each SPG edge, applying C2 to identify undeclared local-to-external flows. Finally, it produces two results: whether description--implementation inconsistency exists, and whether the skill exhibits coarser description. Throughout the process, the taxonomy constrains the LLM's scope: it may only use labels defined in the taxonomy and must not infer capabilities unstated in the description. We set \texttt{temperature=0} for deterministic output and make a single model call per skill. If code-side evidence is insufficient, the LLM returns an explicit uncertain status rather than a forced decision.

\paragraph{Output Schema}
For each skill, the LLM returns a structured JSON result rather than a single label. The output contains: (1) code-side evidence validation indicating whether the SPG evidence is sufficient for reliable checking; (2) declared semantics extracted from \texttt{SKILL.md}; (3) node-level results listing undeclared behavior candidates with their checking outcomes; (4) flow-level results listing undeclared flow candidates with their checking outcomes; (5) summary statistics covering the number of relevant nodes, flows, and flagged mismatches; (6) the final result on whether description--implementation inconsistency exists; (7) the coarser-description result; and (8) a cause summary for later manual review.

\begin{figure}[t]
\centering
\begin{agentpanel}{Prompt template for description--implementation consistency checking}

\textbf{System Role}\\
You are a security auditor for \textbf{programmatic skills}. Check whether the \textbf{description} in \texttt{SKILL.md} is consistent with the \textbf{implementation} represented by a code-side security property graph. Take \texttt{SKILL.md} as the reference boundary.

\paneldivider

\textbf{Input}\\
\textbf{Taxonomy:} allowed security property labels and semantic scope. \textbf{SKILL.md:} the full description of the skill. \textbf{code\_graph\_json:} code-side security property nodes and edges.

\paneldivider

\textbf{Audit Tasks}\\
Node consistency; flow consistency; overall consistency result.

\paneldivider

\textbf{Core Principles}\\
Use only the taxonomy, \texttt{SKILL.md}, and \texttt{code\_graph\_json}. Do not infer unstated capabilities, flows, or labels. Return \texttt{graph\_extraction\_uncertain} if evidence is insufficient.

\paneldivider

\textbf{Output}\\
Strict JSON containing declared semantics, node-level and flow-level results, and the two final checking results.
\end{agentpanel}
\caption{Prompt template for description--implementation consistency checking. It takes the taxonomy, \texttt{SKILL.md}, and the code-side SPG as input, and returns structured node-level, flow-level, and overall checking results.}
\label{fig:consistency_checking_prompt}
\end{figure}

\section{Evaluation}
\label{sec:evaluation}

\subsection{Experimental Setup}
\label{subsec:setup}

We evaluate \textsc{SkillScope} on the full evaluation dataset of 4,556 programmatic skills, measuring how well its automatic classifications agree with human judgment. The 4,556 \texttt{SKILL.md} files carry no pre-existing inconsistency labels; the ground truth used in this section is produced by the human review described next.

To produce that ground truth, two authors independently review all 4,556 skills. The reviewers apply the same C1/C2 conditions and taxonomy as \textsc{SkillScope}, ensuring a like-for-like comparison. The two checkers use the same description-side reference and taxonomy, but differ in their implementation-side input. \textsc{SkillScope} takes \texttt{SKILL.md}, the SPG, and the taxonomy as inputs, whereas each reviewer examines \texttt{SKILL.md}, the raw implementation files, and the same taxonomy. The only difference is the representation of the implementation evidence: \textsc{SkillScope} sees a serialized SPG, whereas reviewers read the raw source files directly.

We call this protocol \emph{double-blind human review} in two senses: each reviewer is blinded to the other reviewer's judgment, and both reviewers are blinded to \textsc{SkillScope}'s outputs. Disagreements are resolved through discussion, with unresolved cases referred to a designated adjudicator. The checking stage uses GPT-5 with a fixed prompt template and \texttt{temperature=0}, making a single model call per skill.

For ablation and baseline experiments, we use a random sample of 300 skills from the 4,556-skill dataset. The size of 300 balances statistical coverage with API cost: we evaluate four configurations on this subset (the full system, two ablations, and one baseline), each invoking one LLM call per skill, so the total call count grows with both the subset size and the number of configurations. The same human-review protocol identifies 38 of these 300 as confirmed inconsistencies, which we treat as the ground truth on this subset; all configurations share these 300 skills and labels.

For cross-model comparison, we use a smaller, deliberately constructed 30-skill subset that is stratified across our four collection sources and oversamples positive cases. The 9 inconsistencies among the 30 are confirmed by the same human-review protocol applied to the full dataset, rather than pre-selected by any automated criterion. Two factors motivate this design. First, cross-model evaluation re-runs every skill under each candidate LLM, so the per-skill API cost is multiplied by the number of models compared. Second, under the natural inconsistency prevalence on the full dataset (about 9.4\%), a random 30-skill sample would contain only 2--3 inconsistencies on average, too few to distinguish models on inconsistency recall.

\paragraph{Metric}
For description--implementation inconsistency, we use skill-level precision and recall:
\(
\text{Precision} = TP/(TP+FP),\;
\text{Recall} = TP/(TP+FN),
\)
where \(TP\) is the number of skills flagged by \textsc{SkillScope} and confirmed by manual review as true inconsistency, \(FP\) is the number of flagged skills not confirmed as true inconsistency, and \(FN\) is the number of true inconsistency cases missed by \textsc{SkillScope}.

\paragraph{Compute and Human Effort}
All local computation (CPG export via Joern, SPG node localization, and BFS edge discovery) was performed on a single machine running Ubuntu 22.04 LTS with an Intel Core Ultra 9 285K processor (24 cores), 46\,GB RAM, and 1\,TB storage. No GPU was used. On the full dataset of 4,556 skills, Joern CPG export took approximately 4 hours and SPG construction (node localization and BFS edge discovery) took approximately 2 hours. LLM-based consistency checking was performed via API calls to GPT-5 (main experiments), Gemini-2.5-flash, and Llama-3.3-70b (cross-model comparison); no local model inference was required. The GPT-5 checking stage for the full dataset (4,556 single-call invocations) took approximately 10 days, dominated by API rate limits and per-call latency rather than local computation. The human review of all 4,556 skills required approximately 9 working days per reviewer (two reviewers in parallel, each independently inspecting every skill).

\subsection{Detection Accuracy}
\label{subsec:detection_accuracy}

Table~\ref{tab:eval_overall_results} shows the detection results on the full dataset. Among the 4,556 skills, \textsc{SkillScope} flagged 487 as inconsistent. Of these, 413 were confirmed by human review (TP) and 74 were false positives. Human review also identified 15 false negatives, resulting in a precision of 84.8\% and a recall of 96.5\%.

Table~\ref{tab:confusion_matrix} shows the full three-class confusion matrix. All 1,106 skills classified as coarser description were confirmed by human review, with none reclassified as inconsistency. The asymmetry follows from the two decision boundaries: coarser description is a containment question (whether an implementation detail falls within an already-declared capability), which admits a clear yes/no answer; inconsistency is a boundary-crossing question, where judging whether a declared capability is broad enough to cover a specific detail admits more variation and accounts for the 74 false positives.

\paragraph{False-Positive Patterns}
Table~\ref{tab:rq1_fp_patterns} groups the 74 false positives into five patterns. The dominant error mode (FP1, 35 cases) is that \textsc{SkillScope} treats finer-grained implementation detail as inconsistency even when the relevant higher-level behavior is already covered by the description; for example, a specific file path is flagged as an undeclared object even though the description states ``read and process local files.'' FP2 (23 cases) reflects situations where the description covers the platform/API/secret direction but does not itemize concrete flow-level or storage details. FP3 (12 cases) covers engineering, setup, observability, or execution choices treated as boundary expansion. FP4 and FP5 together account for 4 cases involving runtime evidence beyond \texttt{SKILL.md} or insufficient code-side evidence.

\paragraph{False-Negative Patterns}
The 15 false negatives fall into two patterns. FN1 (14 cases) stems from coverage limitations in the SPG construction stage: the keyword-based localization rules match direct call patterns such as \texttt{requests.post()} and \texttt{os.getenv()}, but may miss equivalent operations expressed through SDK client objects, custom helper functions, framework-specific integration logic, or configuration construction whose call signatures do not match the predefined patterns. For example, an SDK call like \texttt{client.chat.completions.create(...)}, which carries the configured API key to an external service, does not match the keyword rules and is therefore not localized as an SPG node. FN2 (1 case) reflects an input-presentation difference: API-access context can be more readily observed by reading the full source holistically than by traversing the more fragmented graph form, which we treat as a presentation issue rather than a graph-construction defect.

\subsection{Component Contribution}
\label{subsec:component_contribution}
Table~\ref{tab:ablation_results} shows ablation and baseline results on the 300-skill subset. \textsc{SkillScope} (full) achieves an F1 of 91.1\% on this subset, consistent with the 90.3\% on the full dataset (Table~\ref{tab:eval_overall_results}).

Removing the taxonomy (w/o taxonomy) substantially increases false positives. FP rises from 5 to 13, reducing precision from 87.8\% to 72.3\%. Without the taxonomy's scope constraint, the LLM treats finer implementation details as inconsistencies, including cases that should instead be classified as coarser description.

Removing the SPG (w/o SPG), where the LLM receives the raw source code instead, increases false negatives (FN rises from 2 to 8), reducing recall from 94.7\% to 79.0\%. Analysis of the missed cases shows that most involve cross-node flow relationships. Raw source code contains substantial non-security logic that can obscure security-relevant operations, and longer inputs may also be subject to context-window compression. The SPG retains only security-relevant operations and their flows, providing a cleaner signal. The precision of w/o SPG (90.9\%) is slightly higher than \textsc{SkillScope} (full), because the cases that raw-code mode detects tend to be the most prominent inconsistencies.

The baseline (code + SKILL.md, without either the SPG or the taxonomy) shows the lowest F1 of 73.0\%, confirming that both components contribute to overall performance.

Beyond the pipeline components, the choice of LLM backend can also affect results. To quantify this, we re-run the full \textsc{SkillScope} pipeline on the 30-skill cross-model subset under three different LLMs. Table~\ref{tab:model_comparison} shows results with alternative LLMs. GPT-5 achieves the highest F1 (90.0\%) and is used for all main experiments, followed by Gemini-2.5-flash (85.7\%) and Llama-3.3-70b (72.7\%). GPT-5 and Gemini-2.5-flash both achieve 100\% recall on this subset; their F1 gap is driven entirely by precision. Llama-3.3-70b has both lower precision and lower recall, with its additional false positives concentrated in cases where finer implementation details are misclassified as boundary violations.

\subsection{Prevalence and Patterns}
\label{subsec:prevalence_patterns}
Across the 4,556 skills, human review confirmed 428 cases as inconsistent, giving a population prevalence of 9.4\%. Of these, \textsc{SkillScope} correctly captured 413 (the remaining 15 are characterized as false negatives in Section~\ref{subsec:detection_accuracy}). Table~\ref{tab:rq2_inconsistency_patterns} groups the 413 \textsc{SkillScope}-captured cases into six patterns. The dominant pattern (IC1, 212 cases, 51.3\%) is undeclared credential usage and forwarding to external services: the description may state a local task such as ``analyze code style,'' or describe the intended use of an external service, but does not cover that credentials will be read and then sent to that service. The implementation contains operations such as reading an API key, constructing an authorization header, or attaching a token to an outgoing request. These cases combine undeclared behavior (credential access) with an undeclared flow (local data to an external service). This is followed by undeclared flows from user input or local content to execution or system commands (IC3, 91 cases, 22.0\%), and undeclared local execution, permission changes, or dependency-related capability expansion (IC6, 54 cases, 13.1\%). These results show that confirmed inconsistency cases often involve both undeclared behaviors and undeclared flows rather than isolated node-level differences.

A further 1,106 skills (24.3\%) were classified as coarser description. Table~\ref{tab:rq3_coarse_fine_patterns} shows their breakdown. The dominant patterns are more specific local file handling (LU1, 419 cases) and minor implementation-side details (LU2, 321 cases) such as logging, caching, or input validation that fall within an already-covered task scope. A representative coarser-description pattern is local state, cache, or configuration handling: the implementation includes state reads, cache access, configuration-file handling, or session-related operations that serve the declared capability without expanding the security boundary. These cases confirm that a substantial portion of description--implementation differences reflect granularity mismatch rather than security-boundary expansion.

Taken together, the inconsistency and coarser-description breakdowns point to a clear asymmetry: granularity-mismatch cases concentrate on local-only details that stay within the declared task scope, whereas confirmed inconsistencies are dominated by undeclared flows that move local credentials, user input, or local content toward external destinations or local execution (IC1, IC3, IC6 jointly account for 357/413, 86.4\%).

\begin{table}[!t]
  \caption{Inconsistency detection results on the full evaluation dataset (4,556 skills).}
  \label{tab:eval_overall_results}
  \centering
  \small
  \setlength{\tabcolsep}{5pt}
  \renewcommand{\arraystretch}{1.05}
  \begin{tabular}{lcccccccc}
    \toprule
    Flagged & TP & FP & FN & Precision & Recall & F1 \\
    \midrule
    487 & 413 & 74 & 15 & 84.8\% & 96.5\% & 90.3\% \\
    \bottomrule
  \end{tabular}
\end{table}

\begin{table}[t]
  \caption{Confusion matrix on the full evaluation dataset.}
  \label{tab:confusion_matrix}
  \centering
  \small
  \setlength{\tabcolsep}{5pt}
  \begin{tabular}{lccc}
    \toprule
    & \multicolumn{3}{c}{Human review} \\
    \cmidrule(lr){2-4}
    System output & Inconsistency & Coarser desc. & Consistent \\
    \midrule
    Inconsistency (487) & 413 & 0 & 74 \\
    Coarser desc. (1,106) & 0 & 1,106 & 0 \\
    Consistent (2,963) & 15 & 0 & 2,948 \\
    \bottomrule
  \end{tabular}
\end{table}

\begin{table*}[!t]
  \caption{Ablation and baseline results on a 300-skill random subset (38 true inconsistency cases). \emph{Baseline} feeds the LLM with raw source code and \texttt{SKILL.md}, without either the SPG or the taxonomy.}
  \label{tab:ablation_results}
  \centering
  \footnotesize
  \renewcommand{\arraystretch}{1.05}
  \begin{tabular*}{\textwidth}{@{\extracolsep{\fill}}lrrrrrrr@{}}
    \toprule
    Configuration & Flagged & TP & FP & FN & Prec. & Recall & F1 \\
    \midrule
    \textsc{SkillScope} (full) & 41 & 36 & 5 & 2 & 87.8\% & 94.7\% & 91.1\% \\
    \quad w/o taxonomy & 47 & 34 & 13 & 4 & 72.3\% & 89.5\% & 80.0\% \\
    \quad w/o SPG & 33 & 30 & 3 & 8 & 90.9\% & 79.0\% & 84.5\% \\
    \midrule
    Baseline & 47 & 31 & 16 & 7 & 66.0\% & 81.6\% & 73.0\% \\
    \bottomrule
  \end{tabular*}
\end{table*}

\begin{table*}[t]
  \caption{Cross-model comparison on a 30-skill subset with oversampled positive cases.}
  \label{tab:model_comparison}
  \centering
  \small
  \renewcommand{\arraystretch}{1.05}
  \begin{tabular*}{\textwidth}{@{\extracolsep{\fill}}lrrrrrrr@{}}
    \toprule
    Model & Flagged & TP & FP & FN & Prec. & Recall & F1 \\
    \midrule
    GPT-5 & 11 & 9 & 2 & 0 & 81.8\% & 100\% & 90.0\% \\
    Gemini-2.5-flash & 12 & 9 & 3 & 0 & 75.0\% & 100\% & 85.7\% \\
    Llama-3.3-70b & 13 & 8 & 5 & 1 & 61.5\% & 88.9\% & 72.7\% \\
    \bottomrule
  \end{tabular*}
\end{table*}

\begin{table}[!t]
  \caption{Main false-positive patterns in the full evaluation dataset.}
  \label{tab:rq1_fp_patterns}
  \centering
  \footnotesize
  \setlength{\tabcolsep}{3pt}
  \renewcommand{\arraystretch}{1.05}
  \begin{tabular}{@{}l>{\raggedright\arraybackslash}p{0.74\columnwidth}r@{}}
    \toprule
    ID & Pattern & Count \\
    \midrule
    FP1 & Declared capability already covers local I/O, persistence, or state/config handling. & 35 \\
    FP2 & Declaration covers the platform/API/secret direction; concrete flow-level or storage details not itemized. & 23 \\
    FP3 & Engineering, setup, observability, or execution details are implementation choices, not boundary expansion. & 12 \\
    FP4 & Relevant declaration evidence exists at runtime beyond \texttt{SKILL.md}. & 1 \\
    FP5 & Available evidence insufficient for a full inconsistency check. & 3 \\
    \midrule
    \multicolumn{2}{@{}l}{Total} & 74 \\
    \bottomrule
  \end{tabular}
\end{table}

\begin{table}[!t]
  \caption{Confirmed inconsistency patterns in the full evaluation dataset.}
  \label{tab:rq2_inconsistency_patterns}
  \centering
  \footnotesize
  \setlength{\tabcolsep}{3pt}
  \renewcommand{\arraystretch}{1.05}
  \begin{tabular}{@{}l>{\raggedright\arraybackslash}p{0.46\columnwidth}r>{\raggedright\arraybackslash}p{0.22\columnwidth}@{}}
    \toprule
    ID & Pattern & Count & Examples \\
    \midrule
    IC1 & Undeclared credential usage and forwarding to external services. & 212 & docs-seeker, antigravity-quota \\
    IC2 & Undeclared local persistence, config reads/writes, or state writes. & 21 & sequential-thinking, rag-query \\
    IC3 & Undeclared flows from user input or local content to execution or system commands. & 91 & clean-code-reviewer, torch-compile \\
    IC4 & Undeclared external platform integration or data-destination changes. & 19 & secret-scanner, theme-gen \\
    IC5 & Code behavior directly exceeding the described security boundary. & 16 & playwright-skill, clipit \\
    IC6 & Undeclared local execution, permission changes, or dependency expansion. & 54 & kotlin-in-action, create-design-board \\
    \midrule
    \multicolumn{2}{@{}l}{Total} & 413 & \\
    \bottomrule
  \end{tabular}
\end{table}

\begin{table}[!t]
  \caption{Main patterns for coarser-description cases in the full evaluation dataset.}
  \label{tab:rq3_coarse_fine_patterns}
  \centering
  \footnotesize
  \setlength{\tabcolsep}{3pt}
  \renewcommand{\arraystretch}{1.05}
  \begin{tabular}{@{}l>{\raggedright\arraybackslash}p{0.40\columnwidth}r>{\raggedright\arraybackslash}p{0.28\columnwidth}@{}}
    \toprule
    ID & Pattern & Count & Reason \\
    \midrule
    LU1 & Code adds specific file paths, reports, or persistence locations under declared file handling. & 419 & Details serve the declared capability; no undeclared exfiltration or execution. \\
    LU2 & Broad alignment; code contains only minor implementation-side details. & 321 & Differences are minor and do not support an inconsistency conclusion. \\
    LU3 & Code adds logging, observability, or debugging details not itemized in description. & 135 & Observability details remain within the declared task scope. \\
    LU4 & Code adds cache, state, config, or session handling under a declared capability. & 108 & Local state objects support the declared capability rather than extend it. \\
    LU5 & Code introduces limited additional nodes (e.g., health checks, input validation). & 85 & No dangerous flows; limited to small implementation-side details. \\
    LU6 & Code adds setup, directory creation, or packaging steps under a declared workflow. & 38 & Workflow details rather than new capabilities. \\
    \midrule
    \multicolumn{2}{@{}l}{Total} & 1,106 & \\
    \bottomrule
  \end{tabular}
\end{table}

\section{Discussion}
\label{sec:discussion}

\paragraph{Implications}
For skill developers, descriptions need not enumerate every implementation detail, but should clearly state the security-relevant capabilities, resources, and local-to-external flows that define the skill's security boundary. This distinction matters in practice: implementation details may remain within an already declared capability without constituting a security-boundary violation.

For platform providers, automated consistency checking can support review prioritization by surfacing skills with potential security-boundary violations. Distinguishing inconsistency from coarser description allows platforms to route high-priority inconsistency cases to human reviewers while treating coarser-description cases as informational. Because \textsc{SkillScope} requires only a single LLM call per skill after graph construction, the checking cost scales linearly with the number of skills, making it practical to integrate into skill submission or update pipelines.

For end users, the checking results can serve as a transparency signal that surfaces the gap between what the description promises and what the implementation does, supporting informed decisions before installing or invoking a skill. Host LLMs that mediate skill invocation can consume the same signal at runtime to refuse or warn on flagged skills.

\paragraph{Limitations}
The SPG construction relies on static localization rules, which may miss security-relevant behaviors embedded in SDKs, dynamically loaded code, or unconventional patterns. This pattern accounts for 14 of the 15 false negatives reported in Section~\ref{subsec:detection_accuracy}, and motivates either extending the rule set or replacing keyword matching with a learning-based classifier over CPG nodes as future work. The current taxonomy covers the security properties observed in the analyzed dataset but may require extension as skill ecosystems evolve. The consistency checking depends on a single LLM call per skill; multi-round or ensemble-based approaches may improve robustness but were not explored.

\paragraph{Threats to Validity}
The reported metrics depend on the model used (GPT-5) and may vary under other models, as the cross-model comparison in Table~\ref{tab:model_comparison} suggests. Human review involves inherent subjectivity, particularly at the boundary between inconsistency and coarser description; we mitigate this through independent double-blind review with discussion-based conflict resolution and adjudication for unresolved cases. Reported recall is an upper bound conditional on SPG coverage: behaviors that the extraction rules do not localize never reach the checking stage and are counted in neither TP nor FP. The dataset is drawn from four sources (Anthropic official skills, OpenAI, SkillsMP, and skills.rest) and may not represent all programmatic skill ecosystems. Code-side SPG construction inherits the language coverage of Joern's frontend (Python, JavaScript, TypeScript, and Go in our implementation); skills in other languages are excluded, which may limit generalizability to ecosystems with different language distributions. The analysis assumes non-adversarial skill authors; intentional obfuscation or evasion of consistency checking is not addressed.

\paragraph{Reproducibility}
We provide an anonymized implementation of \textsc{SkillScope} at \url{https://anonymous.4open.science/r/SkillScope-4D66/}. The repository contains the full source code, the dataset collection and filtering scripts, the taxonomy and reference annotation set, the SPG construction and consistency-checking pipeline, and evaluation scripts for reproducing the main experimental results reported in Section~\ref{sec:evaluation}. The supplementary README includes environment configuration, data-access instructions, and step-by-step commands for reproducing detection accuracy, ablation, and cross-model comparison experiments. The repository will be de-anonymized upon acceptance.

\section{Related Work}
\label{sec:relatedwork}

\paragraph{Security Risks in Extensible Software Ecosystems.}
Studies on package and extension ecosystems show that third-party components can introduce behaviors not visible from their exposed interface~\cite{zimmermann2019small,duan2020towards}. Programmatic skills share this characteristic: users rely on the description, while the implementation determines runtime behavior. Our work focuses specifically on the consistency between these two layers.

\paragraph{Security of LLM Agent Ecosystems.}
Recent work examines security risks in LLM agent ecosystems, including empirical studies of skill-level risks~\cite{liu2026agent,liu2026malicious}, attacks via malicious skill files or hidden instructions~\cite{schmotz2026skill,wang2026skills}, and MCP-level threats such as tool poisoning~\cite{hou2025model,wang2026mcptox}. Defenses include security auditing and runtime protection~\cite{zhang2026agent,hu2025agentsentinel}. These studies focus on identifying risks or attack vectors. Our work addresses a different question: whether a skill's description is consistent with the security-relevant behaviors in its implementation.

\paragraph{LLMs for Program Analysis.}
LLMs are increasingly applied to vulnerability detection and repair~\cite{wang2025contemporary,zhou2025large}, though prompting on raw code alone does not reliably outperform static analysis~\cite{ceka2024can,gnieciak2025large}. Combining LLMs with static-analysis outputs has shown improvements~\cite{li2024iris,du2025minimizing,li2025automated}. Our setting differs: we do not detect code vulnerabilities, but assess whether a natural-language description is consistent with the security behaviors in the implementation. This requires cross-modal reasoning over text and structured code-side evidence.

\paragraph{Specification--Implementation Consistency.}
Detecting inconsistencies between natural-language descriptions and code has been studied in software engineering. Wen et al.~\cite{wen2019large} conduct a large-scale empirical study of code-comment inconsistencies. Panthaplackel et al.~\cite{panthaplackel2021deep} propose deep-learning models for just-in-time detection of comment-code inconsistency. Liu et al.~\cite{liu2023docchecker} apply code LLMs to detect and resolve code-comment mismatches. These works operate at the function or statement level, comparing inline comments or docstrings with adjacent code. Our setting differs in two respects: the specification is a standalone skill-level description (\texttt{SKILL.md}) rather than an inline comment, and the consistency question is specifically about security-relevant behaviors rather than general semantic alignment.

\section{Conclusion}
\label{sec:conclusion}
We study description--implementation inconsistency in programmatic skills by asking whether the security-relevant behaviors in the implementation remain within the description's declared scope. We construct a security property taxonomy through manual analysis of 920 skills and propose \textsc{SkillScope}, which combines source-level security property graph construction with LLM-assisted consistency checking. \textsc{SkillScope} outputs both whether inconsistency exists and whether the skill exhibits coarser description.

Our evaluation on 4,556 skills with double-blind human review shows that \textsc{SkillScope} achieves a precision of 84.8\% and a recall of 96.5\%, with confirmed inconsistency affecting 9.4\% of skills and coarser description accounting for 24.3\%, confirming the necessity of distinguishing granularity mismatch from security-boundary expansion. Ablation experiments and cross-model comparisons demonstrate the contribution of both the SPG and the taxonomy. Future work includes extending the taxonomy to cover emerging skill behaviors, exploring ensemble-based checking, and studying adversarial robustness against intentional evasion.

\bibliographystyle{IEEEtran}
\bibliography{xguard}

@inproceedings{yamaguchi2014modeling,
  title={Modeling and discovering vulnerabilities with code property graphs},
  author={Yamaguchi, Fabian and Golde, Nico and Arp, Daniel and Rieck, Konrad},
  booktitle={2014 IEEE symposium on security and privacy},
  pages={590--604},
  year={2014},
  organization={IEEE}
}

@misc{joern_docs,
  author       = {{Joern}},
  title        = {Overview | Joern Documentation},
  year         = {2026},
  howpublished = {\url{https://docs.joern.io/}},
  note         = {Accessed: 2026-04-14}
}

@misc{anthropic_skills,
  author       = {{Anthropic}},
  title        = {Extend Claude with Skills},
  year         = {2026},
  howpublished = {\url{https://docs.anthropic.com/en/docs/claude-code/skills}},
  note         = {Accessed: 2026-04-14}
}

@misc{skillsmp_home,
  author       = {{SkillsMP}},
  title        = {SkillsMP: Agent Skills Marketplace},
  year         = {2026},
  howpublished = {\url{https://skillsmp.com/}},
  note         = {Accessed: 2026-04-14}
}

@misc{skillsrest_home,
  author       = {{skills.rest}},
  title        = {Agent Skills Library},
  year         = {2026},
  howpublished = {\url{https://skills.rest/}},
  note         = {Accessed: 2026-04-14}
}

@misc{openai_skills,
  author  = {{OpenAI}},
  title   = {Skills in ChatGPT},
  year    = {2026},
  url     = {https://help.openai.com/en/articles/20001066-skills-in-chatgpt},
  urldate = {2026-04-14},
  note    = {Official documentation}
}

@misc{anthropic_overview,
  author       = {{Anthropic}},
  title        = {Claude Code Overview},
  year         = {2026},
  howpublished = {\url{https://docs.anthropic.com/en/docs/agents-and-tools/claude-code/overview}},
  note         = {Claude Code Docs. Accessed: 2026-04-14}
}

@misc{anthropic_security,
  author       = {{Anthropic}},
  title        = {Security},
  year         = {2026},
  howpublished = {\url{https://docs.anthropic.com/en/docs/claude-code/security}},
  note         = {Claude Code Docs. Accessed: 2026-04-14}
}

@misc{github_create_skills,
  author       = {{GitHub}},
  title        = {Creating Agent Skills for GitHub Copilot},
  year         = {2026},
  howpublished = {\url{https://docs.github.com/en/copilot/how-tos/use-copilot-agents/cloud-agent/create-skills}},
  note         = {GitHub Docs. Accessed: 2026-04-14}
}

@misc{github_about_skills,
  author       = {{GitHub}},
  title        = {About Agent Skills},
  year         = {2026},
  howpublished = {\url{https://docs.github.com/en/copilot/concepts/agents/about-agent-skills}},
  note         = {GitHub Docs. Accessed: 2026-04-14}
}

@misc{github_add_skills,
  author       = {{GitHub}},
  title        = {Adding Agent Skills for GitHub Copilot CLI},
  year         = {2026},
  howpublished = {\url{https://docs.github.com/en/copilot/how-tos/copilot-cli/customize-copilot/add-skills}},
  note         = {GitHub Docs. Accessed: 2026-04-14}
}

@inproceedings{zimmermann2019small,
  title={Small world with high risks: A study of security threats in the npm ecosystem},
  author={Zimmermann, Markus and Staicu, Cristian-Alexandru and Tenny, Cam and Pradel, Michael},
  booktitle={28th USENIX Security symposium (USENIX security 19)},
  pages={995--1010},
  year={2019}
}

@article{duan2020towards,
  title={Towards measuring supply chain attacks on package managers for interpreted languages},
  author={Duan, Ruian and Alrawi, Omar and Kasturi, Ranjita Pai and Elder, Ryan and Saltaformaggio, Brendan and Lee, Wenke},
  journal={arXiv preprint arXiv:2002.01139},
  year={2020}
}

@article{schmotz2026skill,
  title={Skill-Inject: Measuring Agent Vulnerability to Skill File Attacks},
  author={Schmotz, David and Beurer-Kellner, Luca and Abdelnabi, Sahar and Andriushchenko, Maksym},
  journal={arXiv preprint arXiv:2602.20156},
  year={2026}
}

@article{liu2026malicious,
  title={Malicious agent skills in the wild: A large-scale security empirical study},
  author={Liu, Yi and Chen, Zhihao and Zhang, Yanjun and Deng, Gelei and Li, Yuekang and Ning, Jianting and Zhang, Ying and Zhang, Leo Yu},
  journal={arXiv preprint arXiv:2602.06547},
  year={2026}
}

@article{liu2026agent,
  title={Agent Skills in the Wild: An Empirical Study of Security Vulnerabilities at Scale},
  author={Liu, Yi and Wang, Weizhe and Feng, Ruitao and Zhang, Yao and Xu, Guangquan and Deng, Gelei and Li, Yuekang and Zhang, Leo},
  journal={arXiv preprint arXiv:2601.10338},
  year={2026}
}

@article{wang2026skills,
  title={When Skills Lie: Hidden-Comment Injection in LLM Agents},
  author={Wang, Qianli and Ma, Boyang and Xu, Minghui and Zhang, Yue},
  journal={arXiv preprint arXiv:2602.10498},
  year={2026}
}

@article{hou2025model,
  author = {Hou, Xinyi and Zhao, Yanjie and Wang, Shenao and Wang, Haoyu},
  title = {Model Context Protocol (MCP): Landscape, Security Threats, and Future Research Directions},
  journal = {ACM Transactions on Software Engineering and Methodology},
  year = {2026},
  doi = {10.1145/3796519},
  url = {https://doi.org/10.1145/3796519},
  note = {Just Accepted}
}

@inproceedings{wang2026mcptox,
  title={MCPTox: A Benchmark for Tool Poisoning on Real-World MCP Servers},
  author={Wang, Zhiqiang and Gao, Yichao and Wang, Yanting and Liu, Suyuan and Sun, Haifeng and Cheng, Haoran and Shi, Guanquan and Du, Haohua and Li, Xiangyang},
  booktitle={Proceedings of the AAAI Conference on Artificial Intelligence},
  volume={40},
  number={42},
  pages={35811--35819},
  year={2026},
  doi={10.1609/aaai.v40i42.40895},
  url={https://doi.org/10.1609/aaai.v40i42.40895}
}

@article{zhang2026agent,
  title={Agent Audit: A Security Analysis System for LLM Agent Applications},
  author={Zhang, Haiyue and Nian, Yi and Zhao, Yue},
  journal={arXiv preprint arXiv:2603.22853},
  year={2026}
}

@inproceedings{hu2025agentsentinel,
  title={AgentSentinel: An End-to-End and Real-Time Security Defense Framework for Computer-Use Agents},
  author={Hu, Haitao and Chen, Peng and Zhao, Yanpeng and Chen, Yuqi},
  booktitle={Proceedings of the 2025 ACM SIGSAC Conference on Computer and Communications Security},
  pages={3535--3549},
  year={2025},
  doi={10.1145/3719027.3765064},
  url={https://doi.org/10.1145/3719027.3765064}
}

@article{wang2025contemporary,
  title={A Contemporary Survey of Large Language Model Assisted Program Analysis},
  author={Wang, Jiayimei and Ni, Tao and Lee, Wei-Bin and Zhao, Qingchuan},
  journal={Transactions on Artificial Intelligence},
  volume={1},
  number={1},
  pages={105--129},
  year={2025},
  doi={10.53941/tai.2025.100006},
  url={https://doi.org/10.53941/tai.2025.100006}
}

@article{zhou2025large,
  title={Large Language Model for Vulnerability Detection and Repair: Literature Review and the Road Ahead},
  author={Zhou, Xin and Cao, Sicong and Sun, Xiaobing and Lo, David},
  journal={ACM Transactions on Software Engineering and Methodology},
  volume={34},
  number={5},
  pages={145:1--145:31},
  year={2025},
  doi={10.1145/3708522},
  url={https://doi.org/10.1145/3708522}
}

@article{ceka2024can,
  title={Can llm prompting serve as a proxy for static analysis in vulnerability detection},
  author={Ceka, Ira and Qiao, Feitong and Dey, Anik and Valecha, Aastha and Kaiser, Gail and Ray, Baishakhi},
  journal={arXiv preprint arXiv:2412.12039},
  year={2024}
}

@article{gnieciak2025large,
  title={Large language models versus static code analysis tools: A systematic benchmark for vulnerability detection},
  author={Gnieciak, Damian and Szandala, Tomasz},
  journal={IEEE Access},
  volume={13},
  pages={198410--198422},
  year={2025},
  publisher={IEEE}
}

@inproceedings{li2024iris,
  title={IRIS: LLM-Assisted Static Analysis for Detecting Security Vulnerabilities},
  author={Li, Ziyang and Dutta, Saikat and Naik, Mayur},
  booktitle={The Thirteenth International Conference on Learning Representations},
  year={2025},
  note={ICLR 2025 Poster},
  url={https://openreview.net/forum?id=9LdJDU7E91}
}

@article{du2025minimizing,
  title={Minimizing False Positives in Static Bug Detection via LLM-Enhanced Path Feasibility Analysis},
  author={Du, Xueying and Yu, Kai and Wang, Chong and Zou, Yi and Deng, Wentai and Ou, Zuoyu and Peng, Xin and Zhang, Lingming and Lou, Yiling},
  journal={arXiv preprint arXiv:2506.10322},
  year={2025}
}

@article{li2025automated,
  title={Automated static vulnerability detection via a holistic neuro-symbolic approach},
  author={Li, Penghui and Yao, Songchen and Korich, Josef Sarfati and Luo, Changhua and Yu, Jianjia and Cao, Yinzhi and Yang, Junfeng},
  journal={arXiv preprint arXiv:2504.16057},
  year={2025}
}

@inproceedings{panthaplackel2021deep,
  title={Deep Just-In-Time Inconsistency Detection Between Comments and Source Code},
  author={Panthaplackel, Sheena and Li, Junyi Jessy and Gligoric, Milos and Mooney, Raymond J.},
  booktitle={Proceedings of the AAAI Conference on Artificial Intelligence},
  volume={35},
  number={1},
  pages={427--435},
  year={2021}
}

@inproceedings{wen2019large,
  title={A Large-Scale Empirical Study on Code-Comment Inconsistencies},
  author={Wen, Fengcai and Nagy, Csaba and Bavota, Gabriele and Lanza, Michele},
  booktitle={Proceedings of the 27th International Conference on Program Comprehension (ICPC)},
  pages={53--64},
  year={2019},
  organization={IEEE}
}

@inproceedings{liu2023docchecker,
  title={DocChecker: Bootstrapping Code Large Language Model for Detecting and Resolving Code-Comment Inconsistencies},
  author={Liu, Hao and Wang, Yanlin and Cai, Zhao and Zhu, Ping and Zan, Daoguang and Cui, Yijiang and Shi, Bei and Ma, Yongji},
  booktitle={Proceedings of the 46th International Conference on Software Engineering: Companion Proceedings (ICSE-Companion)},
  pages={114--118},
  year={2024}
}

\end{document}